\documentclass[english,12pt]{article}
\usepackage[cp1251]{inputenc}
\usepackage{babel}
\usepackage{amsfonts, amsmath}
\newcommand{\be}{\begin{equation}} \newcommand{\ee}{\end{equation}}

\begin{document}
\title{Density Matrix and Dynamical aspects of Quantum Mechanics
with Fundamental Length} \thispagestyle{empty}

\author{A.E.Shalyt-Margolin\thanks
{Phone (+375) 172 883438; e-mail:a.shalyt@mail.ru; alexm@hep.by
},$\;$J.G.Suarez
\thanks{Phone (+375) 172 883438
; e-mail: suarez@hep.by; jsuarez@tut.by} }
\date{}
\maketitle
 \vspace{-25pt}
{\footnotesize\noindent National Centre of Particles and High
Energy Physics, Bogdanovich Str.153, Minsk 220040, Belarus\\}

\vspace{0.25cm}
 {\ttfamily{\footnotesize

 }}

\rm\normalsize \vspace{0.25cm}
\begin{abstract}
In this paper Quantum Mechanics with Fundamental Length is built
as a deformation of Quantum Mechanics. To this aim an approach is
used which does not take into account commutator deformation as
usually it is done, but density matrix deformation. The
corresponding deformed density matrix, which is called density
pro-matrix is given explicitly. It properties have been
investigated as well as some dynamical aspects of the theory. In
particular, the deformation of Liouville equation is analyzed in
detail. It was shown that Liouville equation in Quantum Mechanics
appears as a low energy limit of deformed Liouville equation in
Quantum Mechanics with Fundamental Length. Some implications of
obtained results are presented as well as their application to the
calculation of black hole entropy.
 \end{abstract}
\newpage
\section{Introduction}
The last decade quite intensively was investigated Quantum
Mechanics with Fundamental Length (QMFL). The main motivation for
these investigations is the description of quantum gravity effects
which become considerable only at the Planck scale. To research
nature at this scale it is necessary to take into account the
concept of minimal length, as it was shown in \cite{r1}. The
concept of fundamental length was introduced in papers on
 String Theory using General
Uncertainty Relations (GUR) \cite{r2}.

In this paper QMFL is analyzed from the measurement procedure
point of view. It was shown that if the generally accepted
measurement rules are used, then density matrix should be deformed
or in other words, it should be changed by it progenetrix (density
pro-matrix) with $Sp[\rho]<1$, which appears when Quantum
Mechanics (QM) is deformed. As deformation parameter was chosen
the quantity $\beta=l_{min}^{2}/x^{2}$, where $l_{min}$ is the
minimal length and $x$ is the scale. This deformation conducts to
QMFL. In this paper that deformation is described explicitly. It
was shown that QM appears at the grained scale limit (low energy
scale). In such a way the paradigm of inflationary model contains
two different (non equivalent) versions of Quantum Mechanics: the
first one describes nature at the Planck scale (QMFL) as well as
the second one is obtained as the limit when we come back from
Planck scale to low energy one (QM). From the given below
arguments we conclude that some well-defined concepts in QM (for
example, pure state, zero entropy and others) appear only in the
low energy limit.

 In this
paper dynamical aspects of QMFL have been analyzed. A prototype of
the Liouville equation  has been obtained. It was shown that
Liouville equation appears in the low energy limit. The
interpretation of established facts is discussed as well as some
implications of obtained results. In particular, for the
information paradox in black holes. Our approach differs from
others since we have considered the deformation of density matrix.
At the same time, in other approaches the deformation of
commutators has been considered. This paper has to be considered
as the logical continuation of \cite{r3}.

\section{Analogue of density Matrix in QMFL}
In the last 15 years a lot of papers were issued in which authors,
using different methods as: String Theory \cite{r2}, Gravitation
\cite{r4}, Quantum Theory of Black Holes \cite{r5} and others
\cite{r6} shown that Heisenberg Uncertainty Relations (UR)
\cite{r7} should be modified. In particular, a high energy
addition has to appear
\begin{equation}\label{U1}
\triangle x\geq\frac{\hbar}{\triangle p}+\alpha
L_{p}^2\frac{\triangle p}{\hbar}.
\end{equation}
\\Where $L_{p}$ is the Planck length:
$L_{p}=\sqrt\frac{G\hbar}{c^3}\simeq1,6\;10^{-35}m$ and
 $\alpha > 0$ is a constant. In paper \cite{r4} it was shown
that this
 constant can be chosen equal to 1. However, here we will use
 $\alpha$ as an arbitrary constant without giving it any concrete value.
 The  inequality (\ref{U1}) is quadratic with respect to
$\triangle p$
\begin{equation}\label{U2}
\alpha L_{p}^2({\triangle p})^2-\hbar \triangle x \triangle p+
\hbar^2 \leq0,
\end{equation}
and from it follows the fundamental length is
\begin{equation}\label{U3}
\triangle x_{min}=2\sqrt\alpha L_{p}.
\end{equation}
Since further we are going to base only on the existence of
fundamental length, it is necessary to point out this fact was
established not only from GUR. For instance, in \cite{r8},
\cite{r9} using an ideal experiment dealing with gravitational
field it was obtained the lower bound on limit length, which was
improved in \cite{r10} without GUR to an estimate of the type
$\sim L_{p}$. Furthermore it is necessary again to remember the
review \cite{r1}, in which it was emphasized that Quantum Gravity,
investigated using different approaches, necessarily conducts to
the concept of fundamental length.
 \\Let's consider in some detail the equation (\ref{U3}).
Squaring it left and right side, we obtain
\begin{equation}\label{U4}
(\overline{\Delta\widehat{X}^{2}})\geq 4\alpha L_{p}^{2 },
\end{equation}
or in terms of density matrix
\begin{equation}\label{U5}
Sp[\rho(\widehat{X}-Sp(\rho \widehat{X})^2)]= Sp[(\rho
\widehat{X}^2)-Sp^2(\rho \widehat{X}) ]\geq 4\alpha L_{p}^{2 }>0.
\end{equation}
where $\widehat{X}$ is the coordinate operator. Expression
(\ref{U5}) gives the measuring rule used in QM. However, in the
case considered here, in comparison with QM, the right part of
(\ref{U5}) cannot be done arbitrarily near to zero since it is
limited by $l^{2}_{min}>0$, where due to GUR $l_{min} \sim L_{p}$.

 Apparently, this may be due to
the fact that QMFL with GUR (\ref{U1}) is unitary non-equivalent
to QM with UR. Actually, in QM the left-hand side of (\ref{U11})
can be chosen arbitrary closed to zero, whereas in QMFL this is
impossible. But if two theories are unitary equivalent then, the
form of their spurs should be retained. Besides, a more important
aspect is contributing to unitary non-equivalence of these two
theories: QMFL contains three fundamental constants (independent
parameters) $G$, $c$ and $\hbar$, whereas QM contains only one:
$\hbar$. Within an inflationary model (see \cite{r9}), QM is the
low-energy limit of QMFL (QMFL turns to QM) for the expansion of
the Universe. In this case, the second term in the right-hand side
of (\ref{U1}) vanishes and GUR turn to UR. A natural way for
studying QMFL is to consider this theory as a deformation of QM,
turning to QM at the low energy limit (during the expansion of the
Universe after the Big Bang). We will consider precisely this
option. However differing from author of papers \cite{r5} and
others, we do not deform commutators, but density matrix, leaving
at the same time the fundamental quantum-mechanical measuring rule
(\ref{U5}) without changes. Here the following question may be
formulated: how should be deformed density matrix conserving
quantum-mechanical measuring rules in order to obtain
self-consistent measuring procedure in QMFL? For answering to the
question we will use the R-procedure. For starting let us to
consider R-procedure both at the Planck's energy scale and at the
low-energy one. At the Planck's scale $a \approx il_{min}$ or $a
\sim iL_{p}$, where $i$ is a small quantity. Further $a$ tends to
infinity and we obtain for density matrix $$Sp[\rho a^{2}]-Sp[\rho
a]Sp[\rho a] \simeq l^{2}_{min}\;\; or\;\; Sp[\rho]-Sp^{2}[\rho]
\simeq l^{2}_{min}/a^{2}.$$

 Therefore:

 \begin{enumerate}
 \item When $a < \infty$, $Sp[\rho] =
Sp[\rho(a)]$ and
 $Sp[\rho]-Sp^{2}[\rho]>0$. Then, \newline $Sp[\rho]<1$
 that corresponds to the QMFL case.
\item When $a = \infty$, $Sp[\rho]$ does not depend on $a$ and
$Sp[\rho]-Sp^{2}[\rho]\rightarrow 0$. Then, $Sp[\rho]=1$ that
corresponds to the QM case.
\end{enumerate}
How should be points 1 and 2 interpreted? How does analysis
above-given agree to the main result from \cite{r26} \footnote
{"... there cannot be any physical state which is a position
eigenstate since a eigenstate would of course have zero
uncertainty in position"}? It is in full agreement. Indeed, when
state-vector reduction (R-procedure) takes place in QM then,
always an eigenstate (value) is chosen exactly. In other words,
the probability is equal to 1. As it was pointed out in the
above-mentioned point 1 the situation changes when we consider
QMFL: it is impossible to measure coordinates exactly since it
never will be absolutely reliable. We obtain in all cases a
probability less than 1 ($Sp[\rho]=p<1$). In other words, any
R-procedure in QMFL leads to an eigenvalue, but only with a
probability less than 1. This probability is as near to 1 as far
the difference between measuring scale $a$ and $l_{min}$ is
growing, or in other words, when the second term in (\ref{U1})
becomes insignificant and we turn to QM. Here there is not a
contradiction with \cite{r26}. In QMFL there are not exact
coordinate eigenstates (values) as well as there are not pure
states. In this paper we do not consider operator properties in
QMFL as it was done in \cite{r26} but density-matrix properties.

 The  properties of density matrix in
QMFL and QM have to be different. The only reasoning in this case
may be as follows: QMFL must differ from QM, but in such a way
that in the low-energy limit a density matrix in QMFL must
coincide with the density matrix in QM. That is to say, QMFL is a
deformation of QM and the parameter of deformation depends on the
measuring scale. This means that in QMFL $\rho=\rho(x)$, where $x$
is the scale, and for $x\rightarrow\infty$  $\rho(x) \rightarrow
\widehat{\rho}$, where $\widehat{\rho}$ is the density matrix in
QM.

Since on the Planck's scale $Sp[\rho]<1$, then for such scales
$\rho=\rho(x)$, where $x$ is the scale, is not a density matrix as
it is generally defined in QM. On Planck's scale we name $\rho(x)$
 "density pro-matrix". As follows from the above, the density
matrix $\widehat{\rho}$ appears in the limit
\begin{equation}\label{U12}
\lim\limits_{x\rightarrow\infty}\rho(x)\rightarrow\widehat{\rho},
\end{equation}
when GUR (\ref{U1}) turn to UR  and QMFL turns to QM.

Thus, on Planck's scale the density matrix is inadequate to obtain
all information about the mean values of operators. A "deformed"
density matrix (or pro-matrix) $\rho(x)$ with $Sp[\rho]<1$ has to
be introduced because a missing part of information $1-Sp[\rho]$
is encoded in the quantity $l^{2}_{min}/a^{2}$, whose specific
weight decreases as the scale $a$ expressed in  units of $l_{min}$
is going up.

\section{QMFL as a deformation of QM}
Here we are going to describe QMFL as a deformation of QM using
the density pro-matrix formalism. In this context density
pro-matrix has to be understood as a deformed density matrix in
QMFL. As fundamental deformation parameter use
$\beta=l^{2}_{min}/x^{2 }$, where $x$ is the scale and
$l_{min}\sim L_{p}$.

\noindent {\bf Definition 1.}

\noindent Any system in QMFL is described by the density
pro-matrix
$\rho(\beta)=\sum_{i}\omega_{i}(\beta)|i><i|$, where
\begin{enumerate}
\item $0<\beta\leq1/4$;
\item The vectors $|i>$ form a full orthonormal system;
\item $\omega_{i}(\beta)\geq 0$ and for all $i$ there is a
finite limit $\lim\limits_{\beta\rightarrow
0}\omega_{i}(\beta)=\omega_{i}$;
\item
$Sp[\rho(\beta)]=\sum_{i}\omega_{i}(\beta)<1$,$\sum_{i}\omega_{i}=1$;
\item For any operator $B$ and any $\beta$ there is a
 mean operator $B$, which depends on  $\beta$:
$<B>_{\beta}=\sum_{i}\omega_{i}(\beta)<i|B|i>$.
\end{enumerate}
At last, in order to match our definition with the result of
section 2 the next condition has to be fulfilled:
\begin{equation}\label{U9}
Sp[\rho(\beta)]-Sp^{2}[\rho(\beta)]\approx\beta,
\end{equation}
from which we can find the meaning of the quantity
$Sp[\rho(\beta)]$, which satisfies the condition of definition:
\begin{equation}\label{U10}
Sp[\rho(\beta)]\approx\frac{1}{2}+\sqrt{\frac{1}{4}-\beta}.
\end{equation}
From point 5) it follows,that $<1>_{\beta}=Sp[\rho(\beta)]$.
Therefore for any scalar quantity $f$ we have $<f>_{\beta}=f
Sp[\rho(\beta)]$. In particular, the mean value
$<[x_{\mu},p_{\nu}]>_{\beta}$ is equal to
\begin{equation}\label{U11}
<[x_{\mu},p_{\nu}]>_{\beta}= i\hbar\delta_{\mu,\nu}
Sp[\rho(\beta)]=i\hbar\delta_{\mu,\nu}(\frac{1}{2}+\sqrt{\frac{1}{4}-\beta})
\end{equation}
We obtain density matrix as the limit
$\lim\limits_{\beta\rightarrow 0}\rho(\beta)=\rho$. It is evident,
that in the limit $\beta\rightarrow 0$ we turn to QM. Here we
would like to verify, that both two cases described above
correspond to the meanings of $\beta$. In the first case when
$\beta$ is near to $\frac{1}{4}$. In the second one when it is
near to zero.
\\From the definitions given above it follows that
$<(j><j)>_{\beta}=\omega_{j}(\beta)$. \\From which the condition
of completeness on $\beta$ is
$<(\sum_{i}|i><i|)>_{\beta}=<1>_{\beta}=Sp[\rho(\beta)]$. The norm
of any vector $|\psi>$, assigned to  $\beta$ can be defined as
\\$<\psi|\psi>_{\beta}=<\psi|(\sum_{i}|i><i|)_{\beta}|\psi>
=<\psi|(1)_{\beta}|\psi>=<\psi|\psi> Sp[\rho(\beta)$, where
$<\psi|\psi>$ is the norm in QM, or in other words when
$\beta\rightarrow 0$. By analogy, for probabilistic interpretation
the same situation takes place in the described theory, but only
changing $\rho$ by $\rho(\beta)$.
\\

\renewcommand{\theenumi}{\Roman{enumi}}
\renewcommand{\labelenumi}{\theenumi.}
\renewcommand{\labelenumii}{\theenumii.}

Some remarks:

\begin{enumerate}
\item The considered above limit covers at the same time
Quantum and Classical Mechanics. Indeed, since $\beta\sim L_{p}^{2
}/x^{2 }=G \hbar/c^3 x^{2 }$, so we obtain:
\begin{enumerate}
\item $(\hbar \neq 0,x\rightarrow
\infty)\Rightarrow(\beta\rightarrow
0)$ for QM;
\item $(\hbar\rightarrow 0,x\rightarrow
\infty)\Rightarrow(\beta\rightarrow
0)$ for Classical Mechanics;
\end{enumerate}
\item In reality the parameter of deformation $\beta$
should take the meaning $0<\beta\leq1$. However, as we can see
from (\ref{U10}), and as it was indicated in the section 2,
$Sp[\rho(\beta)]$ is well defined only for $0<\beta\leq1/4$. Some
troubles can appear only for the point with fundamental length,
since if $x=2l_{min}$,then the problem vanishes. At the very point
with fundamental length $x=l_{min}\sim L_{p}$ there is a
singularity, which is connected with the appearance of the complex
meaning of $Sp[\rho(\beta)]$, or in other words it is connected
with the impossibility of diagonalization  density pro-matrix at
this point over the field of real numbers. For this reason
definition 1 at the initial point does not have any sense.
\item We have to consider the question about solutions
of (\ref{U9}). For instance, one of the solutions of (\ref{U9}),
at least at first order on $\beta$ is
$\rho^{*}(\beta)=\sum_{i}\alpha_{i} exp(-\beta)|i><i|$, where all
$\alpha_{i}>0$ do not depend on $\beta$  and their sum is equal 1.
Indeed, we can easy verify that
\begin{equation}\label{U12}
Sp[\rho^{*}(\beta)]-Sp^{2}[\rho^{*}(\beta)]=\beta+O(\beta^{2})
\end{equation}
Here it is necessary to consider that in momentum's representation
$\beta=p^{2}/p^{2}_{pl}$, where $p_{pl}$ is the Planck momentum.
In the case when the quantity $exp(-\beta)$ is present in the
matrix elements it can dump out the contribution of large momentum
in perturbation theory.
\item It is clear, that in the proposed above description
states, which have a probability equal to 1 (pure state), can
appear only in the limit $\beta\rightarrow 0$, or when all states
$\omega_{i}(\beta)$ except one of them are equal to zero, or when
they tend to zero at this limit.
\item We suppose, that all definitions concerning
density matrix can be transfer to the described above deformation
of Quantum Mechanics (QMFL) changing the density matrix $\rho$ by
the density pro-matrix $\rho(\beta)$ and turning then to the low
energy limit $\beta\rightarrow 0$. In particular, for statistical
entropy we have
\begin{equation}\label{U13}
S_{\beta}=-Sp[\rho(\beta)\ln(\rho(\beta))].
\end{equation}
The quantity $S_{\beta}$, evidently never is equal to zero, since
$\ln\rho(\beta)\neq 0$ and, therefore $S_{\beta}$ may be equal to
zero only at the limit $\beta\rightarrow 0$.
\end{enumerate}

\renewcommand{\theenumi}{\arabic{enumi}}

\begin{enumerate}
\item If we carry out a measurement in a defined scale we cannot
consider a density pro-matrix  with a precision, which exceed some
limit of order $\sim10^{-66+2n}$, where $10^{-n}$ is the scale in
which the measurement is carried out. In most of the known cases
this precision is quite enough for considering density pro-matrix
 density matrix. However, at the Planck scale, where quantum
gravity effects cannot be neglected and energy is of the Planck
order the difference between $\rho(\beta)$ and $\rho$ has to be
considered.
\item At the Planck scale the notion "Wave function of the
Universe", introduced by J.A. Wheeler and B. deWitt \cite{r12}
does not work and in this case quantum gravity effects can be
described only with the help of density pro-matrix $\rho(\beta)$.
\item Since density pro-matrix $\rho(\beta)$ depends on the scale in which
the measurement is carried out, then the evolution of the Universe
within inflationary model paradigm \cite{r11} is not an unitary
process, because, otherwise the probability
$p_{i}=\omega_{i}(\beta)$   would be conserved.
\end{enumerate}

\section{Dynamical aspects of QMFL}
Let's suppose that in QMFL density pro-matrix has the form
$\rho[\beta(t),t]$, or in other words it depends on two
parameters: time $t$ and deformation parameter $\beta$, which also
depends on time $\beta=\beta(t)$. Then we have
\begin{equation}\label{U14}
\rho[\beta(t),t]=\sum_{i}\omega_{i}[\beta(t)]|i(t)><i(t)|.
\end{equation}
We obtain the equation
\begin{equation}\label{U15}
\frac{d\rho[\beta(t),t]}{dt}=\sum_{i}
\frac{d\omega_{i}[\beta(t)]}{dt}|i(t)><i(t)|-i[H,\rho(\beta)]
\end{equation}
This is a prototype of the Liouville equation (deformed Liouville
equation in QMFL).

Let's consider some particular cases.
\begin{enumerate}
\item First we consider the process of time
evolution at low energies, or in other words, when $\beta(0)
\approx 0$, $\beta(t)\approx 0$ and $t \to \infty$. Then it is
clear that $\omega_{i}(\beta)\approx \omega_{i} \approx constant$.
The first term in (\ref{U15}) vanishes and we obtain the Liouville
equation.
\item We obtain also the Liouville equation when we turn from
inflation to big scale. Within the inflationary approach the scale
$a \approx e^{Ht}$, where $H$ is the Hubble's constant and $t$ is
time. Therefore $\beta \approx e^{-2Ht}$ and when $t$ is quite big
$\beta \to 0$. In other words $\omega_{i}[\beta] \to \omega_{i}$,
the first term in (\ref{U15}) vanishes and we again obtain the
Liouville equation.
\item At very early stage of inflationary process or even before it
takes place $\omega_{i}[\beta]$ was not a constant and therefore
the first term in (\ref{U15}) should be taking into account. This
way we obtain a deviation from the Liouville equation.
\item At last let's consider the case when $\beta(0) \approx 0$,
$\beta(t)>0$ when $t \to \infty$. In this case we are going from
low energy scale to high energy one and $\beta(t)$ grows when $t
\to \infty$. In this case all terms in (\ref{U15}) are significant
and we obtain an addition to the Liouville equation in the form
$$\sum\frac{d\omega_{i}[\beta(t)]}{dt}|i(t)><i(t)|.$$ This case
can take place when matter go into a Black Hole and it is moving
in the direction of the singularity (to the Planck scale).
\end{enumerate}

\section{Analysis of the Information Paradox Problem in  Black
Holes}
 Results obtained so far allow us anew to give a meaning to
the problem of information loss in a Black Hole
\cite{r13,r14,r15}, at least for Quantum Black Holes with a big
curvature. Indeed, when we deal with Black Holes of this nature
quantum effects are considerable at the Planck scale and as it was
shown above the initial entropy of matter, absorbed by a Black
Hole at this scale cannot be equal to zero. This is in agreement
with the R.Myers's results: pure quantum states do not form Black
Holes \cite{r16}. Due to this result the problem of information
loss on Black Holes should be reformulated in other way, since in
all papers on information paradox zero entropy at the initial
state is compared with nonzero entropy at the final state.
Moreover it is necessary to note, that last time in some papers
 GUR for Black Holes are considered at the very beginning
\cite{r17}. As a consequence of this approach stable remnants with
mass of the order of Planck mass $M_{Pl}$ appear during the
evaporation of Black Holes. The last affirmation allows us to
conclude that Black Holes should not have evaporated completely.
Results given in \cite{r18} can be applied to the semi-classical
case and are not suitable for considering Quantum Black Holes.

On the other hand, from the results obtained above, at least at
the qualitative level, it can be elucidated the answer to the
question how could be lost information in big Black Holes, which
are formed as the result of a star collapse? Our point of view is
closed to the R.Penrose's one \cite{r19}. He considers (in
opposition to S. Hawking) that information in Black Holes is lost
when matter meets a singularity. In our approach information loss
takes place in the same form. Indeed, near to the horizon of
events an approximately pure state with practically equal to zero
initial entropy $S^{in}=-Sp[\rho\ln(\rho)]$, which corresponds to
$\beta\rightarrow 0$, when approaching to a singularity (in other
words is reaching the Planck scale) gives yet non zero entropy
$S_{\beta}=-Sp[\rho(\beta)\ln(\rho(\beta))]>0$, when $\beta >0$.
Therefore entropy increases and information is lost  at this Black
Hole. We can (at the moment also at the qualitative level)
evaluate entropy of Black Holes. Indeed, starting from density
matrix for a pure state at the "entry" of a Black Hole
$\rho_{in}=\rho_{pure}$ with zero entropy $S^{in}=0$, we obtain at
the singularity in the Black Hole density pro-matrix
$\rho_{out}=\rho(\beta)\approx \rho(1/4)$ with entropy
$S^{out}=S_{1/4}=-Sp[\rho(1/4)\ln(\rho(1/4)]= -1/2 \ln1/2 \approx
0.34657$.
 If we take into account that total entropy of a
Black Hole is proportional to quantum area of surface  A, measured
in Planck units of area $L_{p}^2$ \cite{r20}, we obtain the next
value for Quantum entropy of a Black Hole:
\begin{equation}\label{U16}
S_{black  hole}=0.34657 A
\end{equation}

This value differs from the well-known one given by
Bekenstein-Hawking for Black Hole entropy $S_{black
hole}=\frac{1}{4} A$ \cite{r21}. The last value was obtained in
the semi-classical approximation. At the present moment quantum
corrections to this value are investigated \cite{r22}. Our
approach based on the quantum nature of Black Holes allows to
obtain formula (\ref{U16}) from basic principles. Let us to note
here that in the approaches, used up to now to modify Liouville
equation due to information paradox \cite{r23}, the additional
member appearing in the right side of $d\rho/dt$, where $\rho$ is
density matrix,  has the form $$-\frac{1}{2}\sum_{\alpha \beta
\neq 0} (Q^{\beta}Q^{\alpha}\rho+\rho Q^{\beta}Q^{\alpha}-2
Q^{\alpha}\rho Q^{\beta})$$  where $Q^{\alpha}$
  is a full orthogonal set
of Hermitian matrices with $Q^{0} =1$. In this case either
locality or conservation of energy-momentum  tensor is broken
down. In the offered in this paper approach, the added member in
the deformed Liouville equation  has a more natural and beautiful
form, in our opinion: $$\sum_{i}
\frac{d\omega_{i}[\beta(t)]}{dt}|i(t)><i(t)|.$$ All properties of
QM are conserved in the limit $\beta\rightarrow 0$, in which the
added member vanishes and we obtain Liouville equation.

\section{Conclusion}
Measurement procedure in QM consists of two parts: measurement
rules and measurement operator. In this paper we try to answer to
the following question: if measurement rules (\ref{U5}) have not
been changed, then how is deformed measurement operator when QM is
deformed? It is clear that in this case, since one of the
components of the measurement procedure has been deformed (in
particular density matrix), then the measurement procedure itself
should be changed also. Here it is rightful to formulate the
following question: is it correct to use generally accepted
measurement rules in Quantum gravity? Usually in Quantum Gravity
precisely the generally accepted rules are used \cite{r24},
 but measurement
operator is not deformed. However, according to the given above
arguments, the measurement operator should be deformed.

As it was noted in \cite{r1} all known approaches, used to
substantiate Quantum Gravity one way or another conduct to the
concept of "fundamental length". Furthermore GUR (\ref{U1}) which
also conduct to this concept are well incorporated within the
inflationary model \cite{r25}. Therefore to understand nature at
the planck scale, leaving apart the concept of fundamental length
it seems to be no possible.


\end{document}